\documentclass[usenatbib,a4paper,fleqn]{mnras}

\usepackage[T1]{fontenc}
\usepackage{ae,aecompl}

\usepackage{mathptmx}
\usepackage{graphicx}	% Including figure files
\usepackage{amsmath}	% Advanced maths commands
\usepackage{amssymb}	% Extra maths symbols
\usepackage{multicol}   % Multi-column entries in tables
\usepackage{multirow}   % Multi-row entries in tables
\usepackage{hyperref}
\usepackage{textcomp}
\usepackage{enumerate}
\usepackage{ulem}

\newcommand{\be}{\begin{equation}}
\newcommand{\ee}{\end{equation}}
\newcommand{\bea}{\begin{eqnarray}}
\newcommand{\eea}{\end{eqnarray}}

\newcommand{\revised}[2][red]{#2}
\newcommand{\mm}{\,\hbox{mm}}
\newcommand{\km}{\,\hbox{km}}
\newcommand{\mum}{\,\hbox{\textmu{}m}}
\newcommand{\cm}{\,\hbox{cm}}

\newcommand{\AU}{\,\hbox{AU}}
\newcommand{\g}{\,\hbox{g}}
\newcommand{\s}{\,\hbox{s}}
\newcommand{\pc}{\,\hbox{pc}}

\newcommand{\Myr}{\,\hbox{Myr}}
\newcommand{\Gyr}{\,\hbox{Gyr}}

\newcommand{\Mearth}{\,M_{\oplus}}

\newcommand{\Mjup}{\,M_{\text{Jup}}}
\newcommand{\density}{\g/\cm^3}

\newcommand{\AAp}      {Astron. Astrophys.}

\newcommand{\AJ}       {Astron. J.}

\newcommand{\ApJ}      {Astrophys. J.}
\newcommand{\ApJL}      {Astrophys. J. Lett.}
\newcommand{\ApJS}     {Astrophys. J. Suppl.}

\newcommand{\ARAA}     {Ann. Rev. Astron. Astrophys.}

\newcommand{\BAAS}     {BAAS}

\newcommand{\MNRAS}    {MNRAS}

\newcommand{\PASP}     {PASP}

\title[Scattered disc of HR~8799]{The scattered disc of HR~8799}

\author[Geiler et al.]{
        Fabian Geiler $^{1}$,
        Alexander V. Krivov $^{1}$,
        Mark Booth $^{1}$,
        and
        Torsten L\"ohne $^{1}$
       \\
$^{1}$ Astrophysikalisches Institut und Universit\"atssternwarte, 
Friedrich-Schiller-Universit\"at Jena,
           Schillerg\"a{\ss}chen~2--3, 07745 Jena, Germany
          }

\date{Received \textit{31 August 2018}; accepted \textit{21 November 2018}}

\begin{document}
\pagerange{\pageref{firstpage}--\pageref{lastpage}}
\label{firstpage}
\maketitle
\begin{abstract}
HR~8799 is a young F0-type star with four directly imaged giant planets and two debris belts, one located exterior and another one interior to the region occupied by the planetary orbits. Having an architecture similar to that of our Solar System, but also revealing dissimilarities such as high masses of planets\revised{,} a huge radial extent and a high mass of the outer debris belt, HR~8799 is considered to be a benchmark to test formation and evolution models of planetary systems.
Here we focus on the outer debris ring and its relation to the planets.
We demonstrate that the models of the outer disc, proposed previously to reproduce 
Herschel observations, are inconsistent with the ALMA data, and vice versa.
In an attempt to find a physically motivated model that would agree with both observational sets,
we perform collisional simulations.
We show that a narrow planetesimal belt and a radiation pressure induced dust halo cannot account for the observed radial brightness profiles. A single, wide planetesimal disc does not reproduce the data either.
Instead, we propose a two-population model, comprising a Kuiper-Belt-like structure of a low-eccentricity planetesimal population (``the classical Kuiper Belt'') and a high-eccentricity population of comets (``scattered disc'').
We argue that such a structure of the exo-Kuiper belt of HR~8799 could be explained with planet migration scenarios analogous to those proposed for the Kuiper Belt of the Solar System.
\end{abstract}


%------------------------------------------------------------------
% Contents
%------------------------------------------------------------------

\section{Introduction}

Debris discs, consisting of planetesimal remnants of the protoplanetary disc, are a vital component of the study of planet formation.
Studies have shown that many discs coexist with planets \citep{maldonado-et-al-2012,wyatt-et-al-2012,eiroa-et-al-2013,marshall-et-al-2014,moromartin-et-al-2015,meshkat-et-al-2017}.
One such system is HR~8799, an F0V \citep{gray-et-al-2003} star located $41.29\pm 0.15\pc$ \citep{gaia-et-al-2018} away from Earth with four directly imaged giant planets at projected distances of $27$, $43$, $68$, and $15\AU$ \citep{marois-et-al-2008,marois-et-al-2010}.
Additionally warm dust ($6-10\AU$) and a Kuiper Belt analogue ($>100\AU$) have been detected \citep{sadakane-nishida-1986,su-et-al-2009,hughes-et-al-2011,matthews-et-al-2013b,booth-et-al-2016}.
This combination of objects in one system has made HR~8799 the centre of many dynamical studies and stability analyses (e.g. \citealt{reidemeister-et-al-2009}, \citealt{fabrycky-murrayclay-2009},  \citealt{gozdziewski-migaszewski-2009}, \citealt{marois-et-al-2010}, \citealt{gozdziewski-migaszewski-2014}, \citealt{gozdziewski-migaszewski-2018}).
Not only is the stability of the system of great interest, the formation of these objects is also puzzling.
The two leading models for giant gas planet formation are core accretion \citep{pollack-et-al-1996,kenyon-bromley-2009} and formation through gravitational instability in the protoplanetary disc \citep{boss-1997}.
Neither of these can explain the wide orbits and high masses of these planets \citep{marois-et-al-2010,currie-et-al-2011} without including additional mechanisms such as migration \citep{crida-2009} and planet-planet scattering \citep{chatterjee-et-al-2008}.
A further question is which role the debris disc may have played in the evolution of the system.
It might have circularised the orbits of scattered planets or facilitated migration \citep{moore-et-al-2013}.

An analysis of the entire HR~8799 system is necessary to understand how this planetary system has formed and evolved.
Many attempts were made to better observe, model or explain the structure of the debris disc.
While both \cite{su-et-al-2009} and \cite{matthews-et-al-2013b} set the inner edge of the \revised{outer} disc at roughly $100\AU$, \cite{booth-et-al-2016} found it to be at $145\AU$.
The latter implies a large gap between HR~8799b and the inner edge.
The opening of such a wide gap can be explained with the chaotic zone of a planet which depends on the mass, position \citep{wisdom-1980,duncan-et-al-1989} and eccentricity of the planet \citep{pearce-wyatt-2014}.
In the chaotic zone, mean motion resonances overlap and orbits become unstable, thus particles are removed from the system on short dynamical timescales.
The outermost known planet HR~8799 b can create such a zone up to $110\AU$, which is sufficient for the inner edge inferred by \cite{matthews-et-al-2013b}.
Assuming an eccentricity of at least $0.3$ was held by the planet for a long time, planet b can even clear the system up to $145\AU$ \citep{moromartin-et-al-2010}.
As an alternative way to explain this gap, \cite{booth-et-al-2016} propose a fifth planet positioned at $110\AU$ with a mass of $1.25\Mjup$. If farther out or on an orbit with higher eccentricity, the mass of the planet can also be lower.

The cold belt is found to extend to around 300 AU in the models of  \cite{su-et-al-2009} and \cite{matthews-et-al-2013b}, \revised{who also find emission extending out to ~2000 AU, which they attribute to small grains being blown out by radiation pressure.} In the ALMA data, however, this extended emission is not visible and the cold belt is seen to extend to 450 AU \citep{booth-et-al-2016}. A challenge for both observations is to explain why the disc is so extended and still sufficiently excited. The excitation models of \cite{kenyon-bromley-2008}  need Pluto sized objects to stir the disc, but the formation of these objects at the distance of the cold disc of HR 8799 would take longer than the age of the system. Even assuming smaller objects sufficiently stir the disc, self stirring is still not able to produce destructive collisions in the extended disc  \citep{krivov-booth-2018}. This poses the question of how to explain any sort of destructive collision at these distances. So either this extended disc is still primordial \citep{heng-tremaine-2010, krivov-et-al-2013} or the stirring has to be attributed to another mechanism, such as stirring by planets. 

The previous studies of HR~8799's debris disc, however, \revised{utilized ad hoc distributions of dust and not a collisional model.}
The goal of this work is to find a model which can explain the appearance of HR~8799's outer massive debris disc in both the Herschel/PACS \citep{poglitsch-et-al-2010} and the ALMA \citep{brown-et-al-2004} observations, and how it might be related to the planets.

In section 2 we focus on the system itself and the observations we used.
In section 3 our methods are discussed.
In section 4 we present the modelled configurations.
Section 5 discusses the results and their implications for the formation history of the system.
Finally we present our conclusions in section 6.

\begin{table}
\begin{center}
\caption{Photometry of HR~8799 \label{Tab:1}}
\begin{tabular}{cccc}
\hline
\hline
Photometric Band & Magnitude [mag] & Remarks & Ref.\\
\hline
B      & $6.090 \pm 0.300$ &      &(1)   \\
B      & $6.196          $ &      &(2)   \\
B      & $6.210 \pm 0.010$ &      &(3)   \\
B      & $6.214 \pm 0.009$ &      &(4)   \\
V      & $5.960 \pm 0.010$ &      &(4)   \\
V      & $5.959          $ &      &(2)   \\
V      & $5.960 \pm 0.010$ &      &(3)   \\
I      & $5.690 \pm 0.300$ &      &(1)   \\
J      & $5.383 \pm 0.027$ &      &(5)   \\
H      & $5.280 \pm 0.018$ &      &(5)   \\
$\text{K}_\text{S}$ & $5.240 \pm 0.018$ &      & (5) \\
\hline
Wavelength [$\mum$] & Flux [mJy]        &      &     \\
\hline 
9      & $404.035 \pm 17.808$ & (a) & (6)  \\
12     & $278 \pm 26$          & (b) & (7)  \\
12	   &  $267	\pm	25$        & (b) & (8)  \\
18     & $120.533 \pm 80.276$  & (c) & (6)  \\
23.68  & $86.6 \pm 1.7$        & (d) & (9)  \\
60     & $445 \pm 70$          & (e) & (8)  \\
60     & $450 \pm 71$          & (e) & (7)  \\
60     & $412 \pm 21$          &     & (10)  \\
71.42  & $610 \pm 31$          &     & (9)  \\
70     & $537 \pm 15$          & (f) & (11) \\
90     & $585 \pm 41$          &     & (10)  \\
90     & $488.632 \pm 74.838$ & (g) & (12) \\
100    & $687 \pm 20$   	   & (f) & (11) \\
155.89 & $555 \pm 66$   	   &     & (9)  \\
160    & $570 \pm 50$   	   & (f) & (11) \\
250    & $309 \pm 30$   	   & (f) & (11) \\
350    & $163 \pm30$      	   & (f) & (11) \\
350    & $89 \pm 26$    	   &     & (13) \\
500    & $74 \pm 30$    	   & (f) & (11) \\
850    & $10.3 \pm 1.8$ 	   &     & (14) \\
850    & $17.4 \pm 1.5$ 	   &     & (15) \\

1200   & $4.8 \pm 2.7$  	   &     & (16) \\
\hline
\end{tabular}
\end{center}
\textbf{Remarks:}
(a) color corrected 7000K = 1.184
(b) color corrected 5000K = 1.43
(c) color corrected 7000K = 0.990
(d) calibrated with \cite{brott-hauschildt-2005} model 7400 K
(e) color corrected 50K = 0.91;
(f) BG source subtracted;
(g) color corrected 50K = 0.979\\
\textbf{References:}
(1) The USNO-B1.0 Catalogue \citep{monet-et-al-2003};
(2) NOMAD Catalogue \citep{Zacharias-et-al-2004}, from Tycho-2 Catalogue \citep{Hog-et-al-2000};
(3) The Guide Star Catalogue Version 2.3.2 \citep{lasker-et-al-2008};
(4) The Hipparcos and Tycho Catalogues \citep{hipparcos-1997};
(5) 2MASS All-Sky Catalogue \citep{skrutskie-et-al-2006};
(6) Akari/IRC Mid-Infrared All-Sky Survey Point Source Catalogue \citep{ishihara-et-al-2010};
(7) IRAS Faint Source Catalogue, $|b|>10$ , Version 2.0 \citep{moshir-et-al-1990};
(8) IRAS Catalogue of Point Sources, Version 2.0 \citep{helou-walker-1988};
(9) \citep{su-et-al-2009};
(10) \citep{moor-et-al-2006};
(11) \citep{matthews-et-al-2013b};
(12) Akari/FSI All-Sky Survey Point Source Catalogues \citep{yamamura-et-al-2010};
(13) \citep{patience-et-al-2011};
(14) \citep{williams-andrews-2006};
(15) SONS-Survey \citep{holland-et-al-2017};
(16) \citep{sylvester-et-al-1996}
\end{table}

\begin{table*}
\caption{\revised{Initial p}arameters of some proposed disc components and our preferred model.\label{Tab:2}}
\center
\begin{tabular}{c|c|c|c|c|c||c}
\hline							
\multirow{2}{*}{Parameter} & \revised{Dust Distr.} Model  & \revised{Dust Distr.} Model  &  \multirow{2}{*}{Excited disc}& \multirow{2}{*}{Wide Cold disc} &   \multirow{2}{*}{Preferred Model}\\
	   		 	    & Matthews+14	 &	Booth+16   & 				      &					         & 				\\
\hline
\multicolumn{6}{c}{\textit{1st Population}} \\
\revised{$M~[\Mearth]$    } & \revised{ $ 5.9 $ }            & \revised{ $ 42 $ }   &  -               & \revised{ $ 220 $ }            &  \revised{ $ 134 $}\\
$a~[\AU]$         &	$100-310$	        & $145-430$	  &  -        	     &  $150-440$           &	$140-440$         \\
$e$		          &	  -		            &  -          &  - 	    	     &  \revised{$0 - 0.1$} &	\revised{$0 - 0.1$}\\
$\gamma$          &	$1.0$    	        &	$0.6$	  &  -        	     &  $ 0.6$              &	$1.0$               \\
\multicolumn{6}{c}{\textit{2nd Population}} \\
\revised{$M~[\Mearth]$}     & \revised{$ 0.08 $}            &   -         & \revised{$ 220 $}          &  -                  &  \revised{$ 67 $}\\
$a~[\AU]$	      &	$310-2000$	        &	-		  & $360-440$        &	-	               &  $360-440$ \\
$e$		          &     	-	        &	-		  & $0.5-0.6$        &	-	               &  $0.3-0.5$ \\
$\gamma$	      &	   $1.7$	        &	-		  & $1.0$            &	-	               &  $1.0$     \\
\hline
\end{tabular}\\
\raggedright \textbf{Notes:} $a$ stands for the semimajor axis, $e$ is the eccentricity, and $\gamma$ is the slope of the radial distribution of the optical depth with the form $r^{-\gamma}$. \revised{ $M$ stands for the initial mass of the disc. Since the dust distributions models describe observations, their masses reference the current disc.}
 \end{table*}

\section{Previous Models}

Many observations have already peered into the architecture of HR~8799, giving us a detailed image of the system. The infrared excess was discovered by IRAS \citep{sadakane-nishida-1986, zuckerman-song-2004, rhee-et-al-2007}, a warm inner component was observed by Spitzer/IRS \citep{jura-et-al-2004,chen-et-al-2006,su-et-al-2009} and lastly four giant planets were directly imaged by the Keck and Gemini telescopes \citep{marois-et-al-2008,marois-et-al-2010}.
Here we focus primarily on the observations of \cite{matthews-et-al-2013b} and \cite{booth-et-al-2016}, because they resolved the cold disc, with Herschel/PACS and ALMA respectively, at wavelengths dominated by the cold disc emission.
From the former we used the $70$ and $100\mum$ observations, neglecting the $160\mum$ and the additional SPIRE data, as these were strongly affected by emission from a background cloud and their resolution is poorer.
From the latter we got $1.34\mm$ observations, giving us high resolution images at a much longer wavelength and helping us constrain the location of the bigger grains and their parent bodies, the planetesimals.

In both \cite{matthews-et-al-2013b} and \cite{booth-et-al-2016} \revised{parametric models of the surface brightness distribution were fit to the observed profiles. In order to determine how well the model designed to fit the Herschel data does at fitting the the ALMA data and vice versa, we start by reproducing these models, albeit using physically motivated models. In other words, we distribute the dust according to the parameters given in those models but use realistic grain properties and a size distribution.}
Unless stated otherwise, the populations consisted of $\mum$ to $\km$ sized objects distributed between the inner and outer edge of the disc \revised{following a size distribution of $m^{-1.87}$.}
The material, used throughout this work, is a mixture of astrosilicate ($50\%$, \citealt{draine-2003}), ice \citep{li-greenberg-1998}, and vacuum (each $25\%$) with a density of $\rho=2.0\density$, where vacuum is serving as a substitute for porosity.
We ended up with ``dummy models''\revised{, replicas of the ad hoc dust distributions of the other studies, created with our tools,} that allowed us to extrapolate from the original to other wavelengths.
\revised{We note that, due to the difference in modelling technique, we cannot exactly reproduce the models of the previous models, but our representations are close enough to make qualitative comparisons.}

To create the corresponding images, the distribution of the dust and the material properties were taken into account and the thermal emission calculated. These artificial images were then convolved with a Gaussian Beam corresponding to the point spread functions (PSF) of the instruments. We used normalized elliptical Gaussians with sizes of $5.8^{\prime\prime}\times 5.5^{\prime\prime}$ for the $70\mum$ images, $6.9^{\prime\prime}\times 6.7^{\prime\prime}$ for the $100\mum$ images and $1.3^{\prime\prime}\times 1.7^{\prime\prime}$ for the $1340\mum$ images.
\revised{From these images we calculated the azimuthally averaged radial profiles and compare them to the deprojected radial profiles of the observations.}

\revised{In addition to comparing} our models to \revised{these profiles}, we also compare them to the full SED. All of the photometry for the HR~8799 system available in the literature is presented in Table \ref{Tab:1}. We used the standard calibration system of Johnson, to transform the \textit{B}, \textit{V}, \textit{I} magnitudes into flux values for the spectral energy distribution (SED). The \textit{J}, \textit{H} and $\textit{K}_\text{S}$ magnitudes from 2MASS were transformed with the calibrations of \cite{cohen-et-al-2003}. In order to model the stellar component we use data from the PHOENIX model \citep{brott-hauschildt-2005} fitted by \cite{matthews-et-al-2013b}.

In the following we discuss the extrapolated models of \cite{matthews-et-al-2013b} and \cite{booth-et-al-2016}.

\begin{figure*}
\centering
\includegraphics[width=0.8\textwidth, angle=0]{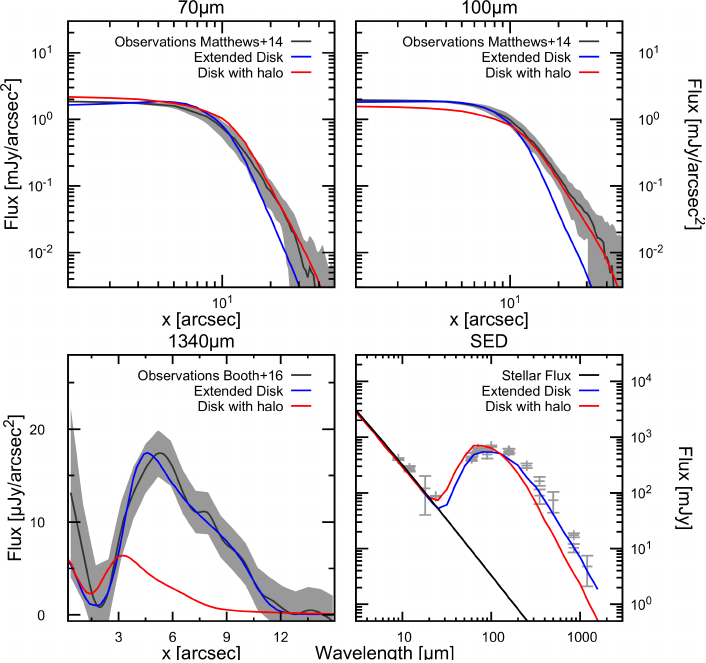}\\
\caption{Radial profiles of the disc at $70, 100$, $1340\mum$ and the SED for the system structures proposed in previous works. In black: observations and error bars (in grey). Red: a debris disc from $100-310\AU$ and a halo of small grains out to $2000\AU$ (as seen in \protect\cite{matthews-et-al-2013b}). Blue: a debris disc with planetesimals from $145\AU$ to $430\AU$ with eccentricities up to $0.1$ (as seen in \protect\cite{booth-et-al-2016}).}
\label{Fig:1}
\end{figure*}

%----------------------------------------------------------------
%------------------ Matthews+14 ---------------------------
%----------------------------------------------------------------
\subsection{Disc with halo\label{Sec:HDisc}}

In the PACS observations the disc extends from $100-310\AU$, but shows a halo stretching to $2000\AU$ (see Tab.~\ref{Tab:2}).
Distances of up to $2000\AU$ are unlikely to be reached by planetesimals, therefore small grains spread by radiation pressure \citep{krivov-et-al-2006} are the most plausible explanation.
Furthermore, \cite{matthews-et-al-2013b} noted differences in the temperatures of the main belt and halo when considering $24\mum+70\mum$ compared to $70\mum+100\mum$ data. This suggests two distinct populations of grains being present.

The slopes of the surface brightness profiles in \cite{matthews-et-al-2013b} differ from wavelength to wavelength, but since we utilize the \revised{geometrical normal} optical depth distributions and not the surface brightness, we \revised{do not need to choose different radial slopes for different wavelengths}.
\revised{Instead we} use the position of the belt, as given in \cite{matthews-et-al-2013b}, but assume a single slope for the optical depth at all wavelengths that best mimicks the slope of the radial profiles (see Tab.~\ref{Tab:2}).
The minimum grain size of the halo is set to the blowout limit of $s_{\text{blow}}\approx 3\mum$ \citep{burns-et-al-1979} while the maximum grain size in the halo is set to three times that value. \revised{Although smaller grains might exist within the halo, their contribution to the radial profiles at wavelengths of $\geq 70\mum$ is negligible.}
One has to point out that in this dummy model (as in the \citealt{su-et-al-2009} and the \citealt{matthews-et-al-2013b} models) the halo is not created by radiation pressure, but consists of small dust grains distributed to resemble a halo.
All other parameters needed are assumed such that our synthetic profiles reproduce the PACS profiles well.
We calculate the error bars as the root mean square of the \revised{azimuthally averaged flux within annuli of a width of $1$ pixel}.

This approach, as seen in red in Fig.~\ref{Fig:1}, does reproduce the outer regions of the Herschel observations with a small flux deviation in the $100\mum$ profile.
In the ALMA profile, however, the synthetic profile peak appears closer to the star while also underpredicting the flux.
The peak location \revised{is} related to the location of the inner edge set by the model.
The lack of flux, however, appears to be an inherent problem with the assumption of a small grain halo, as the SED also follows a much too steep slope at longer wavelengths.
This result differs from the SED in \cite{matthews-et-al-2013b} as they fit the photometry data with a modified black body rather than self consistently fitting the images and the SED.
Larger grains in the halo could remedy the shortcoming, but it is hard to argue that radiation pressure would transport these grains that far out in the system.
So while this offers a solution to the SED shape, the position of the peak in the ALMA profiles remains problematic.
 
\begin{figure*}
\centering
\includegraphics[width=0.8\textwidth, angle=0]{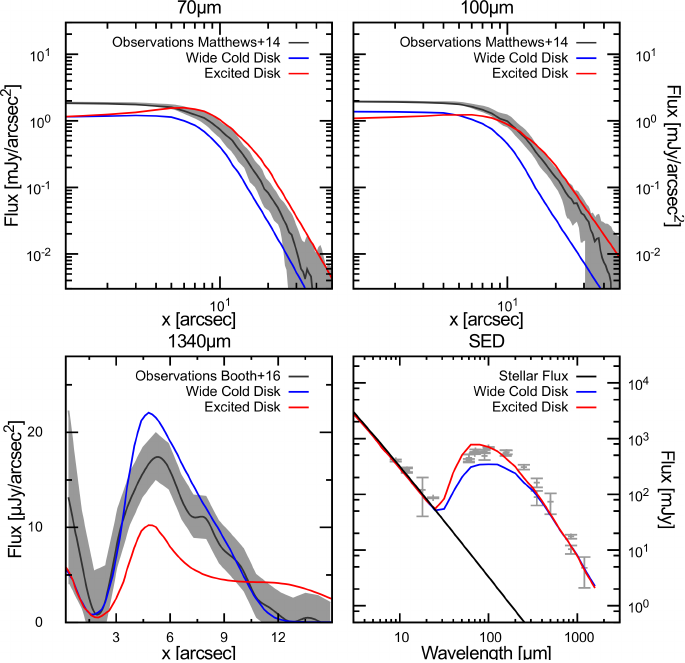}\\
\caption{Profiles and the SED for the wide cold disc and a excited disc model. Black: observations and error bars (in grey). Red: the excited disc model with a belt at $360-440\AU$ and an eccentricity of $0.5-0.6$. Blue: a debris disc with planetesimals from $150\AU$ to $440\AU$ with eccentricities up to $0.1$.}
\label{Fig:2}
\end{figure*}

%---------------------------------------------------------------- 
%------------------ Booth+16 ------------------------------------
%----------------------------------------------------------------
\subsection{Extended disc\label{Sec:ExDisc}}
\cite{booth-et-al-2016} observed the disc around HR~8799 at $1340\mum$ with ALMA in Band 6.
The model they created to fit the observations sets the inner edge of the disc at around $145\AU$ and the outer one at $430\AU$ (see Tab.~\ref{Tab:2}). In the observations themselves the disc appears broad but without halo.
This means the planetesimals are present everywhere across this distance range, as the longer wavelength is more effective at probing larger grains.

Using the parameters given in \cite{booth-et-al-2016} we computed the radial brightness profiles as seen in blue in Fig.~\ref{Fig:1}.
Since the model in \cite{booth-et-al-2016} was used to explain the ALMA observations, we chose any parameters not given in the publication in such a way, that the ALMA observation was well reproduced by our model.
From the resulting dust distribution we calculated  images and the SED (see Fig.~\ref{Fig:1}).
The model is in good agreement with the ALMA observations. Small differences can be explained with the different approach of generating images. The PACS observations, on the other hand, reveal some clear discrepancies. \revised{The profile does not reach out far enough.}

%----------------------------------------------------------------
%------------------ ACE ---------------------------
%----------------------------------------------------------------
\section{\revised{Collisional Evolution Model}\label{Sec:ACE}}
\subsection{Collisional Code}
In order to \revised{construct} a physically motivated model of the disc that is consistent with both Herschel and ALMA wavelengths, we use the ACE code \citep{krivov-et-al-2006, Loehne-et-al-2011, krivov-et-al-2013} to model the long term collisional evolution of a dust distribution.
We assume an azimuthally symmetrical distribution of planetesimals from submicrometer sizes up to $\sim 100\km$.
Using this as a starting point, the \revised{Smoluchowski--Boltzmann} equation is solved to calculate gain and loss of material in this system. 

The code uses the pericentric distance, eccentricity and the mass as phase space variables.
The inclination \revised{and the opening angle are} fixed at half the eccentricity.
\revised{We average the particle densities from the mid-plane to the maximum} inclination.
\revised{If particles enter an unbound orbit, the production rate and dynamical lifetime determine its abundance. The material is considered to have left the system by the next time step and is removed from the simulation.}

\subsection{Collisional physics}
While direct stellar radiation pressure is taken into account, we decided to not consider \revised{Poynting--Robertson} drag, as test runs have shown that it has little impact on the evolution.
Stellar wind drag was also neglected, as F stars such as HR~8799 are not expected to have winds strong enough to influence the disc evolution.
The \revised{composition} was the same as we used for the calculation of the \revised{previous} models, which was mentioned in section \ref{Sec:ExDisc}.
Regarding the material strength, \revised{we adopted the formula of \cite{Loehne-et-al-2011}, which includes the velocity dependence described by \cite{stewart-leinhardt-2009}. The coefficients $b_s=-0.37$, $Q_{D,s}=2\times 10^{-6}$ for the strength regime, $b_g=1.38$, $Q_{D,g}=1\times 10^{-6}$ for the gravity regime, and $v_0 = 3 \km\s^{-1}$ are in accordance with the values found in \cite{benz-asphaug-1999}.
We included disruptive, cratering and bouncing collisions in our runs.
Collisions are considered catastrophic when the impact energy exceeds the specific disruption energy $Q_D^{\ast}$ as described in \cite{stewart-leinhardt-2009} and \cite{Loehne-et-al-2011}.
Cratering of the larger collider occurs when impact energy suffices only to disrupt the smaller collider. For even lower impact energies remnants of both colliders may bounce off each other.
Fragments of all these types of collisions are assumed to follow a mass distribution with number density proportional to $m^{-1.83}$, the average found in \cite{fujiwara-et-al-1977} and \cite{ fujiwara-1986}.}

\subsection{Collisional age of the system}

To model the disc we need its collisional age\revised{, the time since the start of destructive collisions,} which we assume to be roughly equal to the system's age.
HR~8799's age remains a strongly debated topic and values from around $30\Myr$ \citep{zuckerman-et-al-2011,baines-et-al-2012,bell-et-al-2016} up to $1\Gyr$ \citep{moya-et-al-2010} exist \revised{in the literature}.
Most estimates, however, give an age between $30\Myr$ and $100\Myr$ with one line of reasoning being HR~8799's membership of the Columba moving group \citep{torres-et-al-2008, zuckerman-et-al-2011,bell-et-al-2016} which sets the age closer to $30-40\Myr$.
We have checked the probability of HR 8799 being a member of Columba with the Banyan $\Sigma$ code \citep{gagne-et-al-2018} using the latest astrometry from Gaia's 2nd data release \citep{gaia-et-al-2018} and a radial velocity of -12.6$\pm$1.4 km/s \citep{gontcharov-2006}. We find a Columba membership probability of 48.7\%, much lower than that found by previous analyses using pre-Gaia data \citep{zuckerman-et-al-2011, malo-et-al-2013, read-et-al-2018}, casting into doubt the likelihood of membership.
We therefore conclude that the age cannot currently be better determined than $60^{+100}_{-30} \Myr$ \citep{marois-et-al-2008}. \revised{We adopt this age for all further simulations.}

\section{Modelling \revised{and} results}

With simulation runs taking up to several days and a great amount of parameters to vary, choosing the parameters for each run and improving a model is \revised{done on a trial-and-error basis rather than systematically covering a range of parameters}.
As such, the process of finding a well fitting model is always subject to some educated guesses.
This means that, although a well fitting model may be found, other solutions might be possible.

\begin{figure*}
\centering
\includegraphics[width=0.8\textwidth, angle=0]{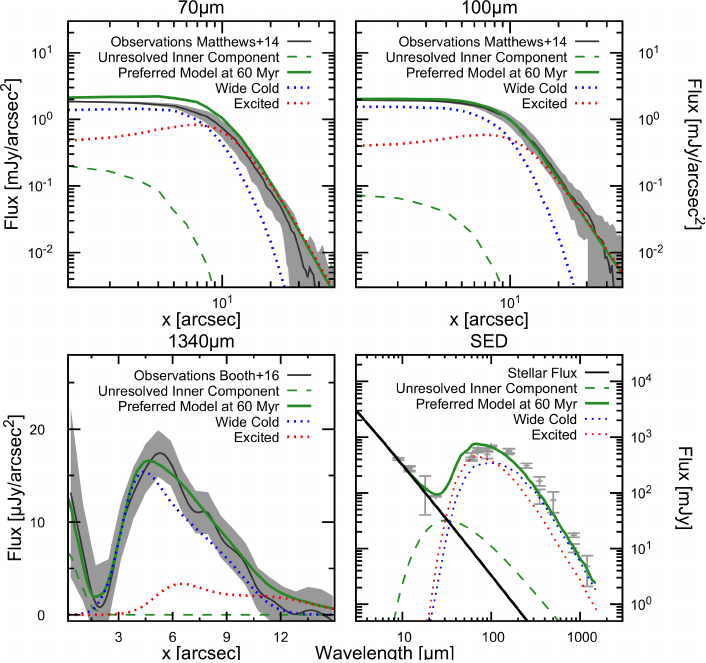}\\
\caption{Profiles and the SED for the preferred scattered disc model. Black: observations and error bars (in grey). Green: The preferred model, comprising \revised{the star, the inner component, } a wide cold disc and a excited one, drawn in a solid line\revised{. The unresolved inner component was additionally} plotted with a dashed line. Red and Blue: The individual populations plotted with a dotted line.}
\label{Fig:3}
\end{figure*}

%----------------------------------------------------------------
%------------------  Excited disc ---------------------
%----------------------------------------------------------------
\subsection{Excited disc}

In the first step we placed the initial planetesimals in orbits with high semimajor axes and high eccentricities, such that the periastra of these orbits lie between $140-220\AU$.
With this we tried to produce a halo of larger grains to address the problems we found with the Herschel/PACS based model.
We varied the eccentricities and semimajor axes in tandem, while trying to reproduce the observations. We again show an example (red lines in Fig.~\ref{Fig:2}) which illustrates the problems with this setup.
Comparison with the PACS data shows that now we find dust far enough out in the system to reproduce the halo mentioned in \cite{matthews-et-al-2013b}, and even overshoot in the $70\mum$ image.
We pay for this by underestimating the profile height significantly further in.
It is even clearer when comparing to the ALMA profile, where the disc is very faint compared to the observation.
This is because the optical depth distribution is much flatter as orbits with high eccentricities distribute the dust over a larger area.
Although collisions occur in the pericentre, more mass is needed for there to be enough mass close to the star.  
The SED on the other hand tells us that we need less material contradicting the other profiles.
Up to this point we did not yet consider the high eccentricities and where they originated.
At a loss to explain the high eccentricities or to solve the mass distribution contradictions we conclude that this model is unable to create the halo and to match the brightness levels of the radial profiles.

%----------------------------------------------------------------
%------------------ Wide Cold disc ---------------------------
%----------------------------------------------------------------

\subsection{Wide cold disc}

In the next step we used a wide cold disc of planetesimals, following the same idea as \cite{booth-et-al-2016}, and evolved it collisionally.
We tried different low values of eccentricity, initial masses and the position of the disc and in the end settled for values only marginally different from those in \cite{booth-et-al-2016}.
\revised{We set the innermost periastra of the initial distribution to be close to the inner edge of the disc at $145\AU$, since that edge was well resolved in the ALMA data.}
Since the \revised{parameters} found here are not unique either, the difference between the two sets of parameters can be neglected.
Just as before, we calculated the radial profiles and SEDs and compared them to the observations in Fig.~\ref{Fig:2}. An example of such a model is seen in blue, with its parameters in Tab.~\ref{Tab:2}. Although this is only an example, its shortcomings are present in all the other similar models we tried.
The PACS images reveal the first major problem: this disc is not \revised{extended} far enough. In both the $70$ and the $100\mum$ image the profile fits nicely in the inner region but falls literally short further out.
It is important to note here that radiation pressure is not effective enough to create the halo seen in the observations. \revised{Neither bound nor unbound grains are abundant enough at these distances.}
This contradicts the previous models of \cite{su-et-al-2009} and \cite{matthews-et-al-2013b}.
The synthetic ALMA image shows a higher peak than the observations, while the SED is too low.
So the profiles at different wavelengths give us different remedies for the model.
We can either increase the total mass to better reproduce the PACS profiles or reduce the total mass to better reproduce the ALMA profiles.
Another problem in this step is that we again have to explain destructive collisions at large distances and as such we need a mechanism of stirring the planetesimals to explain our value of $e=0.1$.
Together with the other problems encountered, this leads us to conclude that the wide cold disc model cannot reproduce the observations.

\subsection{Synthesis model}

We find that a single population is insufficient to reproduce the radial profiles, as the models show that high eccentricities are needed to create the extended emission and low eccentricities needed to reproduce the profile heights.
Catering to both, we tried a synthesis model of both approaches, an excited population and a wide cold population\revised{. The excited population would resemble} a scattered disc. \revised{Explaining the broad disc of HR~8799 with a scattered disc has, in fact, previously been suggested by \cite{wyatt-et-al-2017} who note that this would be a natural result of the proposed fifth planet}.
The parameters were chosen in such a way that in all variations the pericenters of the innermost orbits reach to $130\AU$.
The total masses of these two populations were varied independently, but the same radial slope for the optical depth was assumed.
%------------------Warm Component--------------------------------
%-----------------------------------------------------------------------
Additionally we set a warm component between $6-8\AU$ with a mass needed to adjust the underpredicted $24\mum$ photometry. The contribution of such a warm component to the profiles is negligible except for the innermost part of the ALMA profile. The peak there can be explained by the flux of the warm component and the star.

%----------------------------------------------------------------
%----------------------------------------------------------------
%------------------ Preferred model -----------------------------
%----------------------------------------------------------------
Starting again with the PACS images, we see that this model reproduces the outer slope of the profile while filling the inner regions of the disc.
The preferred model manages to recreate the slope in the outer regions, although in the $70\mum$ image (Fig.~\ref{Fig:3}) we produce a bit too much flux further out.
We plotted the contribution of both populations separately by only calculating the image with objects with $e<0.3$ and another image with objects with $e \geq 0.3$.
The radial profile of these images shows that the excess of flux stems from the overlap of the two populations.
Removing this excess is rather difficult due to the interconnectedness of the different images. Since the error we have in the image is rather small, we are content with the profiles.
By setting the outer edge closer to the star one can reduce the profile height at all PACS wavelengths but also the flux in the slope.
We did not find a configuration of two populations that could solve this problem entirely.
At this point we want to stress, however, that our model for the scattered disc is rather crude and a more nuanced model could eleviate this \revised{(see section \ref{Sec:5.1})}.
The $1340\mum$ profile in Fig.~\ref{Fig:3} also fits the observations rather well.
In summary we see a less severe version of the same problem as in the excited disc model: we overestimate the flux at shorter wavelengths, which is also reflected by the SED.
The model reproduces the profiles to satisfaction, with a few minor discrepancies.

\begin{figure*}
\centering
\includegraphics[width=0.8\textwidth, angle=0]{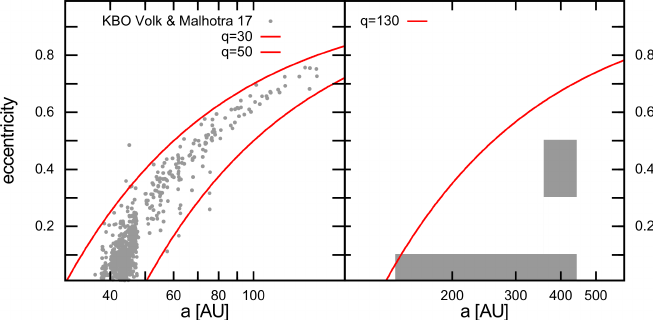}\\
\caption{Left: Kuiper Belt population from \protect\cite{volk-malhotra-2017} in the a-e-plane with rough boundaries. Right: Population of our preferred model with the inner edge of the disc as boundary. This comparison highlights that our model, while similar in approach, is missing the complexity of the continuous distribution of the Kuiper Belt.}
\label{Fig:4}
\end{figure*}

\section{Discussion}

%---------------------------------------------------------------------------------------------
\subsection{Comparison to the Kuiper belt\label{Sec:5.1}}

%---------------------------------------------------------------------------------------------
%------------------ Other Models ---------------------------------------- 
We found that both models of \cite{matthews-et-al-2013b} and \cite{booth-et-al-2016}, while explaining their respective observation, fail once applied to other wavelengths.
The collisional simulations also showed that\revised{, for the parameters tested (which cover a large, although not exhaustive, range), single population discs failed to} reproduce the profiles.
The model most closely reproducing the disc incorporates two populations of parent bodies, with one of them being dynamically excited.
This configuration is roughly similar to the Kuiper Belt of our Solar System with a population of objects of low eccentricity and a scattered population with higher eccentricities (Fig.~\ref{Fig:4}).
%see, e.g. \cite{volk-malhotra-2017}).
The slight differences in the observations and the model can be explained with the restrictions we have with our code.
While in reality the transition from the low eccentricity population to the high eccentricity one is smooth, we are bound by our grid of orbital parameters.
With only two populations used we already almost double the number of the disc parameters in the code. To describe a scattered disc adequately, the amount of parameters would be too unwieldy to obtain any satisfactory solution in any reasonable amount of time.
Although the observed inner edge of the disc serves as a rough approximation of the periastron and restricts the parameter range (see Fig.~\ref{Fig:4}), we still have the masses and many $a$-$e$-configurations to consider.

%---------------------------------------------------------------------------------------------
%------------------ Wilner et al 2018----------------------------------------
\subsection{New SMA data}
In the work of \cite{wilner-et-al-2018} new SMA observations of HR~8799 at $1340\mum$ were published and, in contrast to \cite{booth-et-al-2016}, the visibilities were fitted.
With this fitting technique, the inner edge of the debris disc appears to be at around $110\AU$ with the disc extending to around $500\AU$. This aligns the inner edge more with the Herschel observations.
By moving the cold population closer to the star, we can adjust the inner edge, but the inability to explain the halo with just radiation pressure still persists.
Therefore, \revised{our conclusion that a scattered disc is the best explanation for the observations does not change with a closer inner edge}.

\subsection{Origin of the scattered disc}
The proposed scattered disc can be created in many different ways.
In the Kuiper Belt it is believed to be the result of the migration of the giant planets \citep{gomes-et-al-2017,kaib-sheppard-2016}, and similar evolutionary paths were suggested for the giant planets in HR~8799 \citep{marois-et-al-2010,patience-et-al-2011,dodson-robinson-et-al-2009}.
To discuss its origin, we first look at the masses in the two components in the preferred model and their implications for the protoplanetary disc and planet formation.
%---------------------------------------------------------------------------------------------
%------------------ Mass Estimates and planetformation----------------------------------------
We find an initial mass of $133\Mearth$ for the extended population, and $67\Mearth$ for the scattered population with planetesimals up to the size of $100\km$.
Assuming the scattered population originated as part of the extended population, we calculate the protoplanetary disc mass by assuming that the distribution of the material in the currently observed disc held further-in at the preceding protoplanetary phase.
The resulting total mass of solids is $270\Mearth$, of which $200\Mearth$ are bound in the disc and $70\Mearth$ resided in the region currently occupied by four giant planets.
This mass would have been enough for the planet cores to form, yet the high mass of the scattered population poses the question of the scattering mechanisms.

%----------------------------------------------------------------------------------------------
%----------------------- scattered disc 5 planets    ------------------------------------------
Recently, a fifth planet has been suggested by \citet{read-et-al-2018}
to explain the location of the inner edge of the debris disc as seen in \citet{booth-et-al-2016}.
Modelling its influence on the evolution of the debris disc with N-body simulations, \citet{read-et-al-2018} found a planet with a mass of $0.1\Mjup$ ($\sim 30\Mearth$) and a semi-major axis of $138\AU$ to best explain the shape of the ALMA profile of \citep{booth-et-al-2016}.
Could that planet, if real, be the cause of the scattered disc?
\cite{wyatt-et-al-2017} studied the typical outcomes of close encounters between planets and orbit-crossing planetesimals.
Applying their results to the fifth planet shows that it would scatter most of the orbit-crossing material \revised{onto} bound orbits, thus supporting the fifth planet as the cause of the scattered disc.  
However, this planet's mass of $\sim 30\Mearth$ is much lower than the mass of the scattered population ($\sim 70\Mearth$). It is impossible that a much less massive planet scattered the much more massive planetesimal population.
\cite{read-et-al-2018} furthermore found that planets with a mass in the range from $1\Mjup$ to $0.04\Mjup$ corresponding to either a semimajor axis from $115\AU$ to $130\AU$ or from $140\AU$ to $160\AU$, with the most massive planets being nearest to or farthest away from the star, similarly well explained the observations. Although any planet with a mass in excess of $0.8\Mjup$ would eject orbit-crossing material, a fifth planet capable of scattering $\sim 70\Mearth$ of planetesimals is still possible.

%----------------------------------------------------------------------------------------------
%----------------------- scattered disc 4 planets    ------------------------------------------
As the fifth planet may or may not be real, we shift our focus on to the four confirmed planets.
%------------------------------------Scattering
Referring again to the work by \citet{wyatt-et-al-2017}, one sees that the gas giants of HR~8799, each for itself, would typically eject their orbit-crossing material in unbound orbits.
This material would be on its way out of the system, but, unless this is a continuous process, it should only operate for a relatively short amount of time.
For this process to be continuous, a constant flux of material would need to cross the giant planet's path, which is unlikely.
Therefore we do not consider scattering through the giant planets to be responsible for the scattered disc either.

%------------------------------------Migration
Migration is considered to have occurred in the formation history of HR~8799, as neither core accretion nor gravitational instability could have produced all four planets at their current location \citep{currie-et-al-2011}.
Although we focus in the discussion on outward migration a similar case can be made for inward migration.
Using the formula of \cite{ida-et-al-2000} to calculate the migration rate in our primordial disc, we find a rate of a few 10s of $\AU/\Myr$.
Such a fast migration would allow the planets to quickly reach their current location.
Taking into account that migration stops when the mass within a few Hill radii becomes less than the planet's mass \citep{kirsh-et-al-2009}, we find the fully formed planets in HR~8799 to be too massive to have undergone migration within the protoplanetary disc.
However, if the planets accreted their gas envelope while migrating or even after the migration ceased, then the core masses rather than planet masses would have to be considered.
Migration rate of these, less massive, cores would be lower, and they could easily create a scattered disc of planetesimals instead of ejecting them out of the system.

%------------------------------------PvP - scattering
Planet-planet scattering may have also occurred early on in the history of the HR~8799 system. If the planet cores formed closer to the star by core accretion, such a scattering event could have relocated the planets, bringing them to their current location.
If the planets acquired their gas envelopes after the scattering event, it is possible that the debris disc circularised their orbits, which must have been made eccentric by scattering \citep{bromley-kenyon-2011, currie-et-al-2011}, through dynamical friction of the planet with planetesimals (e.g., \citealt{thommes-et-al-1999}). Obviously it is much easier to circularise the cores than the fully formed gas giants.

\revised{The discussion above hinges on the mass estimates we obtain from our model. It is possible to get larger (smaller) masses by increasing (decreasing) the maximum planetesimal size of $100\km$, while maintain the same collisional evolution. However only a change of an order of magnitude has an impact on the overall mass large enough to change our conclusions. For a less massive disc a fifth planet as proposed in \cite{read-et-al-2018} becomes more plausible. More massive discs slow migration more effectively and one therefore has to consider how far a planet core migrates.}

\subsection{Other systems}
%----------------------------------------Occurence of scattered discs
A natural question is, whether the scattered disc is unique for HR~8799 or can be typical of other debris disc systems.
A distinct feature of the HR~8799 disc is its large radial extent seen in the (sub-)mm images. This disc has a relative width $\Delta r/r\approx 1$, where $\Delta r$ and $r$ are \revised{the} disc's radial extent and mean radius, respectively.
It is natural to expect that potential scattered disc-hosting systems are those that, like HR~8799, appear extended at (sub-)mm wavelengths.
We analysed a list of $26$ debris discs that have been resolved with ALMA or SMA compiled by \cite{matra-et-al-2018}.
One of them is HR~8799 itself.
We inspected the other $25$ systems of the list and found $10$ discs to have a relative width $\Delta r/r > 0.8$.
With $11$ out of $26$ systems being ``extended'' by this criterion, scattered discs could be a common feature. As a caveat, many of these systems have a second dust belt, which impedes a correct determination of the belt width.

As discussed above, formation of scattered discs require planets to be present in the systems. 
Considering the same sample of discs from \cite{matra-et-al-2018},
\cite{krivov-booth-2018} checked which of them do require planets as stirrers to explain the dust production. They found three such discs: HR~8799, HD~95086 and 49~Cet.
While HD~95086 does have one directly imaged planet with a large gap to the inner edge of the disc \citep{rameau-et-al-2013}, planets around 49~Cet are still waiting to be discovered \citep{choquet-et-al-2017}.  
Given the fact that the discs in these two systems are extended and the planets there are either known or at least strongly suspected, we propose them to be particularly good candidates for hosting scattered discs.

\section{Conclusions}
In this work we considered models for the cold debris disc of HR~8799 proposed by \cite{matthews-et-al-2013b} and \cite{booth-et-al-2016}. The former is based on Herschel observations and the latter on ALMA $1.34\mm$ data. We confirmed that both of them reproduced the observation they were based on adequately, but we found them to fail when applied to the other observations. We then searched for a model explaining both sets of observations. To this end, we used the ACE code to collisionally evolve an initial distribution of planetesimals. The resulting dust distribution was then taken to calculate a set of synthesized images to compare to the observations.
Our findings are as follows:

\begin{enumerate}[(i)]
\item Neither a wide planetesimal disc nor an excited narrow disc are able to reproduce the observations, when including radiation pressure, indicating that radiation pressure is not able to reproduce the observed halo.
\item A two-population model with a cold debris disc and a scattered population of planetesimals fits the observations of HR~8799 the best. This architecture is similar to our own Solar System's Kuiper Belt with its classical and scattered components, suggesting its origin to also be similar.
\item A scattered population may be a common feature of debris discs, and probably forms by planets interacting with the disc. For HR~8799 we discussed planet-planet scattering events and migration scenarios involving four to five planets. We believe that the proposed fifth planet can be massive enough to have created the scattered population. The known four planets are too massive to have created a bound scattered population via migration. However, a bound scattered population could have been created by the planet cores through their migration and scattering, if such events occurred before the gas accretion onto the cores.
\end{enumerate}

\section{Acknowledgements}
We thank the anonymous reviewer for their helpful comments and suggestions.
This research was supported by the DFG, grants Kr~2164/13-1, Kr~2164/15-1 and Lo~1715/2-1.
%------------------------------------------------------------------
% Bibliography
%------------------------------------------------------------------

%\input ../0th_Order/journals_short

%--- For compilation with database
%\bibliography{../0th_Order/english}

\begin{thebibliography}{89}
\expandafter\ifx\csname natexlab\endcsname\relax\def\natexlab#1{#1}\fi

\bibitem[{{Baines} {et~al.}(2012){Baines}, {White}, {Huber}, {Jones},
  {Boyajian}, {McAlister}, {ten Brummelaar}, {Turner}, {Sturmann}, {Sturmann},
  {Goldfinger}, {Farrington}, {Riedel}, {Ireland}, {von Braun}, \&
  {Ridgway}}]{baines-et-al-2012}
{Baines}, E.~K., {White}, R.~J., {Huber}, D., {et~al.} 2012, \apj, 761, 57

\bibitem[{Bell {et~al.}(2015)Bell, Mamajek, \& Naylor}]{bell-et-al-2016}
Bell, C. P.~M., Mamajek, E.~E., \& Naylor, T. 2015, Proceedings of the
  International Astronomical Union, 10, 41-48

\bibitem[{{Benz} \& {Asphaug}(1999)}]{benz-asphaug-1999}
{Benz}, W. \& {Asphaug}, E. 1999, \icarus, 142, 5

\bibitem[{{Booth} {et~al.}(2016){Booth}, {Jord{\'a}n}, {Casassus}, {Hales},
  {Dent}, {Faramaz}, {Matr{\`a}}, {Barkats}, {Brahm}, \&
  {Cuadra}}]{booth-et-al-2016}
{Booth}, M., {Jord{\'a}n}, A., {Casassus}, S., {et~al.} 2016, \MNRAS, 460, L10

\bibitem[{{Boss}(1997)}]{boss-1997}
{Boss}, A.~P. 1997, Science, 276, 1836

\bibitem[{{Bromley} \& {Kenyon}(2011)}]{bromley-kenyon-2011}
{Bromley}, B.~C. \& {Kenyon}, S.~J. 2011, \ApJ, 735, 29

\bibitem[{{Brott} \& {Hauschildt}(2005)}]{brott-hauschildt-2005}
{Brott}, I. \& {Hauschildt}, P.~H. 2005, in ESA Special Publication, Vol. 576,
  The Three-Dimensional Universe with Gaia, ed. C.~{Turon}, K.~S. {O'Flaherty},
  \& M.~A.~C. {Perryman}, 565

\bibitem[{{Brown} {et~al.}(2004){Brown}, {Wild}, \&
  {Cunningham}}]{brown-et-al-2004}
{Brown}, R.~L., {Wild}, W., \& {Cunningham}, C. 2004, Advances in Space
  Research, 34, 555

\bibitem[{{Burns} {et~al.}(1979){Burns}, {Lamy}, \& {Soter}}]{burns-et-al-1979}
{Burns}, J.~A., {Lamy}, P.~L., \& {Soter}, S. 1979, \icarus, 40, 1

\bibitem[{{Chatterjee} {et~al.}(2008){Chatterjee}, {Ford}, {Matsumura}, \&
  {Rasio}}]{chatterjee-et-al-2008}
{Chatterjee}, S., {Ford}, E.~B., {Matsumura}, S., \& {Rasio}, F.~A. 2008, \apj,
  686, 580

\bibitem[{{Chen} {et~al.}(2006){Chen}, {Sargent}, {Bohac}, {Kim},
  {Leibensperger}, {Jura}, {Najita}, {Forrest}, {Watson}, {Sloan}, \&
  {Keller}}]{chen-et-al-2006}
{Chen}, C.~H., {Sargent}, B.~A., {Bohac}, C., {et~al.} 2006, \apjs, 166, 351

\bibitem[{{Choquet} {et~al.}(2017){Choquet}, {Milli}, {Wahhaj}, {Soummer},
  {Roberge}, {Augereau}, {Booth}, {Absil}, {Boccaletti}, {Chen}, {Debes}, {del
  Burgo}, {Dent}, {Ertel}, {Girard}, {Gofas-Salas}, {Golimowski}, {G{\'o}mez
  Gonz{\'a}lez}, {Hagan}, {Hibon}, {Hines}, {Kennedy}, {Lagrange}, {Matr{\`a}},
  {Mawet}, {Mouillet}, {N'Diaye}, {Perrin}, {Pinte}, {Pueyo}, {Rajan},
  {Schneider}, {Wolff}, \& {Wyatt}}]{choquet-et-al-2017}
{Choquet}, {\'E}., {Milli}, J., {Wahhaj}, Z., {et~al.} 2017, \apj, 834, L12

\bibitem[{{Cohen} {et~al.}(2003){Cohen}, {Wheaton}, \&
  {Megeath}}]{cohen-et-al-2003}
{Cohen}, M., {Wheaton}, W.~A., \& {Megeath}, S.~T. 2003, \AJ, 126, 1090

\bibitem[{{Crida}(2009)}]{crida-2009}
{Crida}, A. 2009, \ApJ, 698, 606

\bibitem[{{Currie} {et~al.}(2011){Currie}, {Burrows}, {Itoh}, {Matsumura},
  {Fukagawa}, {Apai}, {Madhusudhan}, {Hinz}, {Rodigas}, {Kasper}, {Pyo}, \&
  {Ogino}}]{currie-et-al-2011}
{Currie}, T., {Burrows}, A., {Itoh}, Y., {et~al.} 2011, \apj, 729, 128

\bibitem[{{Dodson-Robinson} {et~al.}(2009){Dodson-Robinson}, {Veras}, {Ford},
  \& {Beichman}}]{dodson-robinson-et-al-2009}
{Dodson-Robinson}, S.~E., {Veras}, D., {Ford}, E.~B., \& {Beichman}, C.~A.
  2009, \apj, 707, 79

\bibitem[{{Draine}(2003)}]{draine-2003}
{Draine}, B.~T. 2003, \ARAA, 41, 241

\bibitem[{{Duncan} {et~al.}(1989){Duncan}, {Quinn}, \&
  {Tremaine}}]{duncan-et-al-1989}
{Duncan}, M., {Quinn}, T., \& {Tremaine}, S. 1989, \icarus, 82, 402

\bibitem[{{Eiroa} {et~al.}(2013){Eiroa}, {Marshall}, {Mora}, {Montesinos},
  {Absil}, {Augereau}, {Bayo}, {Bryden}, {Danchi}, {del Burgo}, {Ertel},
  {Fridlund}, {Heras}, {Krivov}, {Launhardt}, {Liseau}, {L{\"o}hne},
  {Maldonado}, {Pilbratt}, {Roberge}, {Rodmann}, {Sanz-Forcada}, {Solano},
  {Stapelfeldt}, {Th{\'e}bault}, {Wolf}, {Ardila}, {Ar{\'e}valo}, {Beichmann},
  {Faramaz}, {Gonz{\'a}lez-Garc{\'{\i}}a}, {Guti{\'e}rrez}, {Lebreton},
  {Mart{\'{\i}}nez-Arn{\'a}iz}, {Meeus}, {Montes}, {Olofsson}, {Su}, {White},
  {Barrado}, {Fukagawa}, {Gr{\"u}n}, {Kamp}, {Lorente}, {Morbidelli},
  {M{\"u}ller}, {Mutschke}, {Nakagawa}, {Ribas}, \&
  {Walker}}]{eiroa-et-al-2013}
{Eiroa}, C., {Marshall}, J.~P., {Mora}, A., {et~al.} 2013, \AAp, 555, A11

\bibitem[{{Fabrycky} \& {Murray-Clay}(2010)}]{fabrycky-murrayclay-2009}
{Fabrycky}, D.~C. \& {Murray-Clay}, R.~A. 2010, \ApJ, 710, 1408

\bibitem[{{Fujiwara}(1986)}]{fujiwara-1986}
{Fujiwara}, A. 1986, Memorie della Societa Astronomica Italiana, 57, 47

\bibitem[{Fujiwara {et~al.}(1977)Fujiwara, Kamimoto, \&
  Tsukamoto}]{fujiwara-et-al-1977}
Fujiwara, A., Kamimoto, G., \& Tsukamoto, A. 1977, Icarus, 31, 277

\bibitem[{{Gagn{\'e}} {et~al.}(2018){Gagn{\'e}}, {Mamajek}, {Malo}, {Riedel},
  {Rodriguez}, {Lafreni{\`e}re}, {Faherty}, {Roy-Loubier}, {Pueyo}, {Robin}, \&
  {Doyon}}]{gagne-et-al-2018}
{Gagn{\'e}}, J., {Mamajek}, E.~E., {Malo}, L., {et~al.} 2018, \apj, 856, 23

\bibitem[{{Gaia Collaboration} {et~al.}(2018){Gaia Collaboration}, {Brown},
  {Vallenari}, {Prusti}, {de Bruijne}, {Babusiaux}, {Bailer-Jones}, {Biermann},
  {Evans}, {Eyer}, \& et~al.}]{gaia-et-al-2018}
{Gaia Collaboration}, {Brown}, A.~G.~A., {Vallenari}, A., {et~al.} 2018, \aap,
  616, A1

\bibitem[{{Gomes} {et~al.}(2018){Gomes}, {Nesvorn{\'y}}, {Morbidelli},
  {Deienno}, \& {Nogueira}}]{gomes-et-al-2017}
{Gomes}, R., {Nesvorn{\'y}}, D., {Morbidelli}, A., {Deienno}, R., \&
  {Nogueira}, E. 2018, \icarus, 306, 319

\bibitem[{{Gontcharov}(2006)}]{gontcharov-2006}
{Gontcharov}, G.~A. 2006, Astronomy Letters, 32, 759

\bibitem[{{Go{\'z}dziewski} \&
  {Migaszewski}(2009)}]{gozdziewski-migaszewski-2009}
{Go{\'z}dziewski}, K. \& {Migaszewski}, C. 2009, \mnras, 397, L16

\bibitem[{{Go{\'z}dziewski} \&
  {Migaszewski}(2014)}]{gozdziewski-migaszewski-2014}
{Go{\'z}dziewski}, K. \& {Migaszewski}, C. 2014, \mnras, 440, 3140

\bibitem[{{Go{\'z}dziewski} \&
  {Migaszewski}(2018)}]{gozdziewski-migaszewski-2018}
{Go{\'z}dziewski}, K. \& {Migaszewski}, C. 2018, \apjs, 238, 6

\bibitem[{{Gray} {et~al.}(2003){Gray}, {Corbally}, {Garrison}, {McFadden}, \&
  {Robinson}}]{gray-et-al-2003}
{Gray}, R.~O., {Corbally}, C.~J., {Garrison}, R.~F., {McFadden}, M.~T., \&
  {Robinson}, P.~E. 2003, \AJ, 126, 2048

\bibitem[{{Helou} \& {Walker}(1988)}]{helou-walker-1988}
{Helou}, G. \& {Walker}, D.~W., eds. 1988, {Infrared astronomical satellite
  (IRAS) catalogs and atlases. Volume 7: The small scale structure catalog},
  Vol.~7

\bibitem[{{Heng} \& {Tremaine}(2010)}]{heng-tremaine-2010}
{Heng}, K. \& {Tremaine}, S. 2010, \MNRAS, 401, 867

\bibitem[{{H{\o}g} {et~al.}(2000){H{\o}g}, {Fabricius}, {Makarov}, {Urban},
  {Corbin}, {Wycoff}, {Bastian}, {Schwekendiek}, \& {Wicenec}}]{Hog-et-al-2000}
{H{\o}g}, E., {Fabricius}, C., {Makarov}, V.~V., {et~al.} 2000, \AAp, 355, L27

\bibitem[{{Holland} {et~al.}(2017){Holland}, {Matthews}, {Kennedy}, {Greaves},
  {Wyatt}, {Booth}, {Bastien}, {Bryden}, {Butner}, {Chen}, {Chrysostomou},
  {Davies}, {Dent}, {Di Francesco}, {Duch{\^e}ne}, {Gibb}, {Friberg}, {Ivison},
  {Jenness}, {Kavelaars}, {Lawler}, {Lestrade}, {Marshall}, {Moro-Martin},
  {Pani{\'c}}, {Phillips}, {Serjeant}, {Schieven}, {Sibthorpe}, {Vican},
  {Ward-Thompson}, {van der Werf}, {White}, {Wilner}, \&
  {Zuckerman}}]{holland-et-al-2017}
{Holland}, W.~S., {Matthews}, B.~C., {Kennedy}, G.~M., {et~al.} 2017, \mnras,
  470, 3606

\bibitem[{{Hughes} {et~al.}(2011){Hughes}, {Wilner}, {Andrews}, {Williams},
  {Su}, {Murray-Clay}, \& {Qi}}]{hughes-et-al-2011}
{Hughes}, A.~M., {Wilner}, D.~J., {Andrews}, S.~M., {et~al.} 2011, \ApJ, 740,
  38

\bibitem[{{Ida} {et~al.}(2000){Ida}, {Bryden}, {Lin}, \&
  {Tanaka}}]{ida-et-al-2000}
{Ida}, S., {Bryden}, G., {Lin}, D.~N.~C., \& {Tanaka}, H. 2000, \apj, 534, 428

\bibitem[{{Ishihara} {et~al.}(2010){Ishihara}, {Onaka}, {Kataza}, {Salama},
  {Alfageme}, {Cassatella}, {Cox}, {Garc{\'{\i}}a-Lario}, {Stephenson},
  {Cohen}, {Fujishiro}, {Fujiwara}, {Hasegawa}, {Ita}, {Kim}, {Matsuhara},
  {Murakami}, {M{\"u}ller}, {Nakagawa}, {Ohyama}, {Oyabu}, {Pyo}, {Sakon},
  {Shibai}, {Takita}, {Tanab{\'e}}, {Uemizu}, {Ueno}, {Usui}, {Wada},
  {Watarai}, {Yamamura}, \& {Yamauchi}}]{ishihara-et-al-2010}
{Ishihara}, D., {Onaka}, T., {Kataza}, H., {et~al.} 2010, \aap, 514, A1

\bibitem[{{Jura} {et~al.}(2004){Jura}, {Chen}, {Furlan}, {Green}, {Sargent},
  {Forrest}, {Watson}, {Barry}, {Hall}, {Herter}, {Houck}, {Sloan}, {Uchida},
  {D'Alessio}, {Brandl}, {Keller}, {Kemper}, {Morris}, {Najita}, {Calvet},
  {Hartmann}, \& {Myers}}]{jura-et-al-2004}
{Jura}, M., {Chen}, C.~H., {Furlan}, E., {et~al.} 2004, \ApJS, 154, 453

\bibitem[{{Kaib} \& {Sheppard}(2016)}]{kaib-sheppard-2016}
{Kaib}, N.~A. \& {Sheppard}, S.~S. 2016, \aj, 152, 133

\bibitem[{{Kenyon} \& {Bromley}(2008)}]{kenyon-bromley-2008}
{Kenyon}, S.~J. \& {Bromley}, B.~C. 2008, \ApJS, 179, 451

\bibitem[{{Kenyon} \& {Bromley}(2009)}]{kenyon-bromley-2009}
{Kenyon}, S.~J. \& {Bromley}, B.~C. 2009, \ApJ, 690, L140

\bibitem[{{Kirsh} {et~al.}(2009){Kirsh}, {Duncan}, {Brasser}, \&
  {Levison}}]{kirsh-et-al-2009}
{Kirsh}, D.~R., {Duncan}, M., {Brasser}, R., \& {Levison}, H.~F. 2009, \icarus,
  199, 197

\bibitem[{{Krivov} \& {Booth}(2018)}]{krivov-booth-2018}
{Krivov}, A.~V. \& {Booth}, M. 2018, \mnras, 479, 3300

\bibitem[{{Krivov} {et~al.}(2013){Krivov}, {Eiroa}, {L{\"o}hne}, {Marshall},
  {Montesinos}, {del Burgo}, {Absil}, {Ardila}, {Augereau}, {Bayo}, {Bryden},
  {Danchi}, {Ertel}, {Lebreton}, {Liseau}, {Mora}, {Mustill}, {Mutschke},
  {Neuh{\"a}user}, {Pilbratt}, {Roberge}, {Schmidt}, {Stapelfeldt},
  {Th{\'e}bault}, {Vitense}, {White}, \& {Wolf}}]{krivov-et-al-2013}
{Krivov}, A.~V., {Eiroa}, C., {L{\"o}hne}, T., {et~al.} 2013, \ApJ, 772, 32

\bibitem[{{Krivov} {et~al.}(2006){Krivov}, {L{\"o}hne}, \& {Srem{\v
  c}evi{\'c}}}]{krivov-et-al-2006}
{Krivov}, A.~V., {L{\"o}hne}, T., \& {Srem{\v c}evi{\'c}}, M. 2006, \aap, 455,
  509

\bibitem[{{Lasker} {et~al.}(2008){Lasker}, {Lattanzi}, {McLean}, {Bucciarelli},
  {Drimmel}, {Garcia}, {Greene}, {Guglielmetti}, {Hanley}, {Hawkins},
  {Laidler}, {Loomis}, {Meakes}, {Mignani}, {Morbidelli}, {Morrison},
  {Pannunzio}, {Rosenberg}, {Sarasso}, {Smart}, {Spagna}, {Sturch},
  {Volpicelli}, {White}, {Wolfe}, \& {Zacchei}}]{lasker-et-al-2008}
{Lasker}, B.~M., {Lattanzi}, M.~G., {McLean}, B.~J., {et~al.} 2008, \AJ, 136,
  735

\bibitem[{{Li} \& {Greenberg}(1998)}]{li-greenberg-1998}
{Li}, A. \& {Greenberg}, J.~M. 1998, \AAp, 331, 291

\bibitem[{{L{\"o}hne} {et~al.}(2012){L{\"o}hne}, {Augereau}, {Ertel},
  {Marshall}, {Eiroa}, {Mora}, {Absil}, {Stapelfeldt}, {Th{\'e}bault}, {Bayo},
  {Del Burgo}, {Danchi}, {Krivov}, {Lebreton}, {Letawe}, {Magain}, {Maldonado},
  {Montesinos}, {Pilbratt}, {White}, \& {Wolf}}]{Loehne-et-al-2011}
{L{\"o}hne}, T., {Augereau}, J.-C., {Ertel}, S., {et~al.} 2012, \aap, 537, A110

\bibitem[{{Maldonado} {et~al.}(2012){Maldonado}, {Eiroa}, {Villaver},
  {Montesinos}, \& {Mora}}]{maldonado-et-al-2012}
{Maldonado}, J., {Eiroa}, C., {Villaver}, E., {Montesinos}, B., \& {Mora}, A.
  2012, \AAp, 541, A40

\bibitem[{{Malo} {et~al.}(2013){Malo}, {Doyon}, {Lafreni{\`e}re}, {Artigau},
  {Gagn{\'e}}, {Baron}, \& {Riedel}}]{malo-et-al-2013}
{Malo}, L., {Doyon}, R., {Lafreni{\`e}re}, D., {et~al.} 2013, \apj, 762, 88

\bibitem[{{Marois} {et~al.}(2008){Marois}, {Macintosh}, {Barman}, {Zuckerman},
  {Song}, {Patience}, {Lafreni{\`e}re}, \& {Doyon}}]{marois-et-al-2008}
{Marois}, C., {Macintosh}, B., {Barman}, T., {et~al.} 2008, Science, 322, 1348

\bibitem[{{Marois} {et~al.}(2010){Marois}, {Zuckerman}, {Konopacky},
  {Macintosh}, \& {Barman}}]{marois-et-al-2010}
{Marois}, C., {Zuckerman}, B., {Konopacky}, Q.~M., {Macintosh}, B., \&
  {Barman}, T. 2010, Nature, 468, 1080

\bibitem[{{Marshall} {et~al.}(2014){Marshall}, {Moro-Mart{\'{\i}}n}, {Eiroa},
  {Kennedy}, {Mora}, {Sibthorpe}, {Lestrade}, {Maldonado}, {Sanz-Forcada},
  {Wyatt}, {Matthews}, {Horner}, {Montesinos}, {Bryden}, {del Burgo},
  {Greaves}, {Ivison}, {Meeus}, {Olofsson}, {Pilbratt}, \&
  {White}}]{marshall-et-al-2014}
{Marshall}, J.~P., {Moro-Mart{\'{\i}}n}, A., {Eiroa}, C., {et~al.} 2014, \AAp,
  565, A15

\bibitem[{{Matr{\`a}} {et~al.}(2018){Matr{\`a}}, {Marino}, {Kennedy}, {Wyatt},
  {{\"O}berg}, \& {Wilner}}]{matra-et-al-2018}
{Matr{\`a}}, L., {Marino}, S., {Kennedy}, G.~M., {et~al.} 2018, \apj, 859, 72

\bibitem[{{Matthews} {et~al.}(2014){Matthews}, {Kennedy}, {Sibthorpe}, {Booth},
  {Wyatt}, {Broekhoven-Fiene}, {Macintosh}, \& {Marois}}]{matthews-et-al-2013b}
{Matthews}, B., {Kennedy}, G., {Sibthorpe}, B., {et~al.} 2014, \apj, 780, 97

\bibitem[{{Meshkat} {et~al.}(2017){Meshkat}, {Mawet}, {Bryan}, {Hinkley},
  {Bowler}, {Stapelfeldt}, {Batygin}, {Padgett}, {Morales}, {Serabyn},
  {Christiaens}, {Brandt}, \& {Wahhaj}}]{meshkat-et-al-2017}
{Meshkat}, T., {Mawet}, D., {Bryan}, M.~L., {et~al.} 2017, \aj, 154, 245

\bibitem[{{Monet} {et~al.}(2003){Monet}, {Levine}, {Canzian}, {Ables}, {Bird},
  {Dahn}, {Guetter}, {Harris}, {Henden}, {Leggett}, {Levison}, {Luginbuhl},
  {Martini}, {Monet}, {Munn}, {Pier}, {Rhodes}, {Riepe}, {Sell}, {Stone},
  {Vrba}, {Walker}, {Westerhout}, {Brucato}, {Reid}, {Schoening}, {Hartley},
  {Read}, \& {Tritton}}]{monet-et-al-2003}
{Monet}, D.~G., {Levine}, S.~E., {Canzian}, B., {et~al.} 2003, \AJ, 125, 984

\bibitem[{{Mo{\'o}r} {et~al.}(2006){Mo{\'o}r}, {{\'A}brah{\'a}m}, {Derekas},
  {Kiss}, {Kiss}, {Apai}, {Grady}, \& {Henning}}]{moor-et-al-2006}
{Mo{\'o}r}, A., {{\'A}brah{\'a}m}, P., {Derekas}, A., {et~al.} 2006, \ApJ, 644,
  525

\bibitem[{{Moore} {et~al.}(2013){Moore}, {Hasan}, \&
  {Quillen}}]{moore-et-al-2013}
{Moore}, A., {Hasan}, I., \& {Quillen}, A.~C. 2013, \mnras, 432, 1196

\bibitem[{{Moro-Mart{\'{\i}}n} {et~al.}(2010){Moro-Mart{\'{\i}}n}, {Malhotra},
  {Bryden}, {Rieke}, {Su}, {Beichman}, \& {Lawler}}]{moromartin-et-al-2010}
{Moro-Mart{\'{\i}}n}, A., {Malhotra}, R., {Bryden}, G., {et~al.} 2010, \ApJ,
  717, 1123

\bibitem[{{Moro-Mart{\'\i}n} {et~al.}(2015){Moro-Mart{\'\i}n}, {Marshall},
  {Kennedy}, {Sibthorpe}, {Matthews}, {Eiroa}, {Wyatt}, {Lestrade},
  {Maldonado}, {Rodriguez}, {Greaves}, {Montesinos}, {Mora}, {Booth},
  {Duch{\^e}ne}, {Wilner}, \& {Horner}}]{moromartin-et-al-2015}
{Moro-Mart{\'\i}n}, A., {Marshall}, J.~P., {Kennedy}, G., {et~al.} 2015, \apj,
  801, 143

\bibitem[{{Moshir} {et~al.}(1990){Moshir}, Kopan, Conrow, \& {9
  colleagues}}]{moshir-et-al-1990}
{Moshir}, M., Kopan, G., Conrow, T., \& {9 colleagues}. 1990, {IRAS Faint
  Source Catalogue, version 2.0.}

\bibitem[{{Moya} {et~al.}(2010){Moya}, {Amado}, {Barrado}, {Garc{\'\i}a
  Hern{\'a}ndez}, {Aberasturi}, {Montesinos}, \& {Aceituno}}]{moya-et-al-2010}
{Moya}, A., {Amado}, P.~J., {Barrado}, D., {et~al.} 2010, \mnras, 405, L81

\bibitem[{{Patience} {et~al.}(2011){Patience}, {Bulger}, {King}, {Ayliffe},
  {Bate}, {Song}, {Pinte}, {Koda}, {Dowell}, \&
  {Kov{\'a}cs}}]{patience-et-al-2011}
{Patience}, J., {Bulger}, J., {King}, R.~R., {et~al.} 2011, \AAp, 531, L17+

\bibitem[{{Pearce} \& {Wyatt}(2014)}]{pearce-wyatt-2014}
{Pearce}, T.~D. \& {Wyatt}, M.~C. 2014, \mnras, 443, 2541

\bibitem[{{Perryman} {et~al.}(1997){Perryman}, {Lindegren}, {Kovalevsky},
  {Hoeg}, {Bastian}, {Bernacca}, {Cr{\' e}z{\' e}}, {Donati}, {Grenon}, {van
  Leeuwen}, {van der Marel}, {Mignard}, {Murray}, {Le Poole}, {Schrijver},
  {Turon}, {Arenou}, {Froeschl{\' e}}, \& {Petersen}}]{hipparcos-1997}
{Perryman}, M.~A.~C., {Lindegren}, L., {Kovalevsky}, J., {et~al.} 1997, \AAp,
  323, L49

\bibitem[{{Poglitsch} {et~al.}(2010){Poglitsch}, {Waelkens}, {Geis},
  {Feuchtgruber}, {Vandenbussche}, {Rodriguez}, {Krause}, {Renotte}, {van
  Hoof}, {Saraceno}, {Cepa}, {Kerschbaum}, {Agn{\`e}se}, {Ali}, {Altieri},
  {Andreani}, {Augueres}, {Balog}, {Barl}, {Bauer}, {Belbachir}, {Benedettini},
  {Billot}, {Boulade}, {Bischof}, {Blommaert}, {Callut}, {Cara}, {Cerulli},
  {Cesarsky}, {Contursi}, {Creten}, {De Meester}, {Doublier}, {Doumayrou},
  {Duband}, {Exter}, {Genzel}, {Gillis}, {Gr{\"o}zinger}, {Henning},
  {Herreros}, {Huygen}, {Inguscio}, {Jakob}, {Jamar}, {Jean}, {de Jong},
  {Katterloher}, {Kiss}, {Klaas}, {Lemke}, {Lutz}, {Madden}, {Marquet},
  {Martignac}, {Mazy}, {Merken}, {Montfort}, {Morbidelli}, {M{\"u}ller},
  {Nielbock}, {Okumura}, {Orfei}, {Ottensamer}, {Pezzuto}, {Popesso},
  {Putzeys}, {Regibo}, {Reveret}, {Royer}, {Sauvage}, {Schreiber}, {Stegmaier},
  {Schmitt}, {Schubert}, {Sturm}, {Thiel}, {Tofani}, {Vavrek}, {Wetzstein},
  {Wieprecht}, \& {Wiezorrek}}]{poglitsch-et-al-2010}
{Poglitsch}, A., {Waelkens}, C., {Geis}, N., {et~al.} 2010, \AAp, 518, L2

\bibitem[{{Pollack} {et~al.}(1996){Pollack}, {Hubickyj}, {Bodenheimer},
  {Lissauer}, {Podolak}, \& {Greenzweig}}]{pollack-et-al-1996}
{Pollack}, J.~B., {Hubickyj}, O., {Bodenheimer}, P., {et~al.} 1996, \icarus,
  124, 62

\bibitem[{{Rameau} {et~al.}(2013){Rameau}, {Chauvin}, {Lagrange}, {Boccaletti},
  {Quanz}, {Bonnefoy}, {Girard}, {Delorme}, {Desidera}, {Klahr}, {Mordasini},
  {Dumas}, \& {Bonavita}}]{rameau-et-al-2013}
{Rameau}, J., {Chauvin}, G., {Lagrange}, A.-M., {et~al.} 2013, \ApJL, 772, L15

\bibitem[{{Read} {et~al.}(2018){Read}, {Wyatt}, {Marino}, \&
  {Kennedy}}]{read-et-al-2018}
{Read}, M.~J., {Wyatt}, M.~C., {Marino}, S., \& {Kennedy}, G.~M. 2018, \MNRAS

\bibitem[{Reidemeister {et~al.}(2009)Reidemeister, Krivov, Schmidt, Fiedler,
  M\"uller, L"ohne, \& Neuh"auser}]{reidemeister-et-al-2009}
Reidemeister, M., Krivov, A.~V., Schmidt, T. O.~B., {et~al.} 2009, \AAp, 503,
  247

\bibitem[{{Rhee} {et~al.}(2007){Rhee}, {Song}, {Zuckerman}, \&
  {McElwain}}]{rhee-et-al-2007}
{Rhee}, J.~H., {Song}, I., {Zuckerman}, B., \& {McElwain}, M. 2007, \ApJ, 660,
  1556

\bibitem[{{Sadakane} \& {Nishida}(1986)}]{sadakane-nishida-1986}
{Sadakane}, K. \& {Nishida}, M. 1986, \PASP, 98, 685

\bibitem[{{Skrutskie} {et~al.}(2006){Skrutskie}, {Cutri}, {Stiening},
  {Weinberg}, {Schneider}, {Carpenter}, {Beichman}, {Capps}, {Chester},
  {Elias}, {Huchra}, {Liebert}, {Lonsdale}, {Monet}, {Price}, {Seitzer},
  {Jarrett}, {Kirkpatrick}, {Gizis}, {Howard}, {Evans}, {Fowler}, {Fullmer},
  {Hurt}, {Light}, {Kopan}, {Marsh}, {McCallon}, {Tam}, {Van Dyk}, \&
  {Wheelock}}]{skrutskie-et-al-2006}
{Skrutskie}, M.~F., {Cutri}, R.~M., {Stiening}, R., {et~al.} 2006, \AJ, 131,
  1163

\bibitem[{{Stewart} \& {Leinhardt}(2009)}]{stewart-leinhardt-2009}
{Stewart}, S.~T. \& {Leinhardt}, Z.~M. 2009, \ApJL, 691, L133

\bibitem[{{Su} {et~al.}(2009){Su}, {Rieke}, {Stapelfeldt}, {Malhotra},
  {Bryden}, {Smith}, {Misselt}, {Moro-Martin}, \& {Williams}}]{su-et-al-2009}
{Su}, K.~Y.~L., {Rieke}, G.~H., {Stapelfeldt}, K.~R., {et~al.} 2009, \ApJ, 705,
  314

\bibitem[{{Sylvester} {et~al.}(1996){Sylvester}, {Skinner}, {Barlow}, \&
  {Mannings}}]{sylvester-et-al-1996}
{Sylvester}, R.~J., {Skinner}, C.~J., {Barlow}, M.~J., \& {Mannings}, V. 1996,
  \MNRAS, 279, 915

\bibitem[{{Thommes} {et~al.}(1999){Thommes}, {Duncan}, \&
  {Levison}}]{thommes-et-al-1999}
{Thommes}, E.~W., {Duncan}, M.~J., \& {Levison}, H.~F. 1999, Nature, 402, 635

\bibitem[{{Torres} {et~al.}(2008){Torres}, {Quast}, {Melo}, \&
  {Sterzik}}]{torres-et-al-2008}
{Torres}, C.~A.~O., {Quast}, G.~R., {Melo}, C.~H.~F., \& {Sterzik}, M.~F. 2008,
  {Young Nearby Loose Associations}, 757

\bibitem[{{Volk} \& {Malhotra}(2017)}]{volk-malhotra-2017}
{Volk}, K. \& {Malhotra}, R. 2017, \aj, 154, 62

\bibitem[{{Williams} \& {Andrews}(2006)}]{williams-andrews-2006}
{Williams}, J.~P. \& {Andrews}, S.~M. 2006, \ApJ, 653, 1480

\bibitem[{{Wilner} {et~al.}(2018){Wilner}, {MacGregor}, {Andrews}, {Hughes},
  {Matthews}, \& {Su}}]{wilner-et-al-2018}
{Wilner}, D.~J., {MacGregor}, M.~A., {Andrews}, S.~M., {et~al.} 2018, \apj,
  855, 56

\bibitem[{{Wisdom}(1980)}]{wisdom-1980}
{Wisdom}, J. 1980, \AJ, 85, 1122

\bibitem[{{Wyatt} {et~al.}(2017){Wyatt}, {Bonsor}, {Jackson}, {Marino}, \&
  {Shannon}}]{wyatt-et-al-2017}
{Wyatt}, M.~C., {Bonsor}, A., {Jackson}, A.~P., {Marino}, S., \& {Shannon}, A.
  2017, \mnras, 464, 3385

\bibitem[{{Wyatt} {et~al.}(2012){Wyatt}, {Kennedy}, {Sibthorpe}, {Moro-Martin},
  {Lestrade}, {Ivison}, {Matthews}, {Udry}, {Greaves}, {Kalas}, {Lawler}, {Su},
  {Rieke}, {Booth}, {Bryden}, {Horner}, {Kavelaars}, \&
  {Wilner}}]{wyatt-et-al-2012}
{Wyatt}, M.~C., {Kennedy}, G., {Sibthorpe}, B., {et~al.} 2012, \MNRAS, 424,
  1206

\bibitem[{{Yamamura} {et~al.}(2010){Yamamura}, {Makiuti}, {Ikeda}, {Fukuda},
  {Oyabu}, {Koga}, \& {White}}]{yamamura-et-al-2010}
{Yamamura}, I., {Makiuti}, S., {Ikeda}, N., {et~al.} 2010, VizieR Online Data
  Catalog, 2298

\bibitem[{{Zacharias} {et~al.}(2004){Zacharias}, {Monet}, {Levine}, {Urban},
  {Gaume}, \& {Wycoff}}]{Zacharias-et-al-2004}
{Zacharias}, N., {Monet}, D.~G., {Levine}, S.~E., {et~al.} 2004, \BAAS, 36,
  1418

\bibitem[{{Zuckerman} {et~al.}(2011){Zuckerman}, {Rhee}, {Song}, \&
  {Bessell}}]{zuckerman-et-al-2011}
{Zuckerman}, B., {Rhee}, J.~H., {Song}, I., \& {Bessell}, M.~S. 2011, \apj,
  732, 61

\bibitem[{{Zuckerman} \& {Song}(2004)}]{zuckerman-song-2004}
{Zuckerman}, B. \& {Song}, I. 2004, \ApJ, 603, 738

\end{thebibliography}
%\bibliographystyle{aa-package9.0/bibtex/aa}

%--- For compilation without database
%\input paper.bbl.std

\label{lastpage}

\end{document}